\documentclass[letterpaper]{JHEP}
\usepackage[dvips]{graphicx}
\input epsf.tex
\def\be{\begin{equation}}
\def\ee{\end{equation}}
\def\bea{\begin{eqnarray}}
\def\eea{\end{eqnarray}}
\def\ba{\begin{array}}
\def\ea{\end{array}}

\def\part{\partial}

\preprint{SUGP-02/7-1\\
SU-4252-765\\} \keywords{$D$-branes, Myers effect, Non-commutative
geometry}

\title{Obstruction to D-brane Topology Change}

\author{Garnik G. Alexanian\thanks{E-mail: {\tt garnik@physics.syr.edu}}
\hspace{1pt} A.P. Balachandran\thanks{E-mail: {\tt
bal@physics.syr.edu}} and
        Pedro J. Silva\thanks{E-mail: {\tt psilva@physics.syr.edu}}
       \\Physics Department, Syracuse University, Syracuse, New York 13244}
\abstract{A study of the relation between topology change, energy
and Lie algebra representations for fuzzy geometry in connection
to $M$-theory is presented. We encounter two different types of
topology change, related to the different features of the Lie
algebra representations appearing in the matrix models of
$M$-theory. From these studies, we propose a new method of
obtaining non-commutative solutions for the non-Abelian $D$-brane
action found by Myers. This mechanism excludes one of the two
topology changing processes previously found in other
non-commutative solutions of many matrix-based models in
$M$-theory i.e. in M(atrix) theory, Matrix string theory and
non-Abelian $D$-brane physics.}

\begin{document}
\newpage
\section{Introduction}

During the last few years we have seen how non-commutative
geometry has come to play an important role in string theory. It
appears not only at the fundamental Planck distances where a
smooth geometry cannot be trusted, but also at the level of
effective theories  in $D$-brane physics, where the Chan-Paton
factors result in matrix degrees of freedom. In these cases the
effective Lagrangian of $N$ $Dp$-branes comes with built-in
non-commutative features in the form of matrix models. Myers
\cite{mye1} proposed an effective action for these non-Abelian
$D$-branes by demanding consistency with $T$-duality among the
different $Dp$-branes. In particular, he started from the 
well-known $D$9-brane action and proceeded by $T$-dualizing. The
agreement with the weak background actions of Taylor and Van
Raamsdonk \cite{tvr1} was used as a consistency check. Note that
these linearized actions came from a very different theoretical
framework associated with the BFSS M(atrix) theory proposal
\cite{bfss}.

This remarkable characteristic of the built-in non-commutativity
is not new in the framework of $M$-theory. We already have at
least two other examples where this type of construction is found,
namely M(atrix) theory \cite{bfss} and Matrix string theory
\cite{dvv1,mst,sch1,sil1}. In the first case all the
configurations of the theory are to be found in terms of matrix
degrees of freedom, including the fundamental strings and
$D$-branes. In the second case, we have a new formalism in which a
two-dimensional action naturally includes matrix degrees of
freedom representing the `string bits'\footnote{The idea is that
the string can be seen as a chain of partonic degrees of freedom
\cite{tho1}.}, which also incorporate the description of higher
dimensional objects of $M$-theory using non-commutative
configurations.

One of the important properties of this new theoretical framework
(Non-Abelian $Dp$-branes, M(atrix)-theory, Matrix string theory)
lies in the similarity of the mathematical language used to
describe the fundamental objects of $M$-theory, bringing for the 
first time the possibility of describing strings and $D$-branes in a
unified framework, a ``democracy of $p$-branes'' \cite{twn1}.

An essential characteristic of the matrix actions is their
capability to describe non-commutative geometries that correspond
to different extended objects of the theory. For example, higher
dimensional $Dq$-branes may be formed by smaller $Dp$-branes
($q>p$). To be more precise, consider the dielectric effect
\cite{mye1} where $N$ $Dp$-branes form a single $D{(p+2)}$-brane
generating a configuration corresponding to a non-commutative
two-sphere (fuzzy sphere \cite{Madore,Balachandran:2002ig,GPFSP}).
Another kind of construction corresponds to $N$ $D$1-branes
forming $n$ $D$5-branes ($n<N$), this time using a fuzzy
four-sphere \footnote{This particular type of quantum geometry has
been  used extensively, see for example \cite{cas1,tri1,sil1}. }
\cite{mye2}. Also it is worth mentioning that there are other
non-commutative manifolds (apart from the fuzzy $n$-spheres) which
are relevant to matrix models, e.g. fuzzy versions of
 tori, $CP(n)$, $RP(n)$, etc.

In this work we will study the relations between the discrete
partonic picture (the non-commutative picture) and the smooth
geometry that is obtained from it in the limit of large number of
partons, i.e. the ``reconstruction'' of the geometry. Basically,
we would like to understand what is the relation between the
matrix representation of the partonic picture and the resulting
commutative manifold. This relation should be independent of the
$M$-theory object we are using as the fundamental parton once we
invoke the ``$p$-brane democracy idea''.

In particular, we will be working with non-Abelian $D$-brane
actions. In section 2  we will show the constraints that these
classical actions impose on the set of possible algebraic
structures appearing as solutions of the corresponding equations
of motion. Also we will explain the mechanism of obtaining the
fuzzy geometries (quantum geometries) and the reconstruction of
the corresponding commutative manifolds (classical geometries). We
will also discuss how the energy of the different configurations
are crucial to obtain the commutative geometry in the classical
limit. We identify two different types of topology change
occurring in these matrix-valued models. In section 3 we consider
a new mechanism of fixing the geometry by breaking the symmetry
between the solutions of the equations of motion representing
quantum geometries.

The main result is a new class of non-commutative $D$-brane
solutions with obstruction to topology change. It is based on the
important role played by the symmetric representations of the Lie
algebra structure underlying these solutions. We also identify two
types of topology change that can occur in these effective
descriptions.

\section{Quantum vs Classical geometry}

In this section we summarize the 
characteristics that define the non-commutative solutions found in
non-Abelian $D$-brane physics which are importatant for us.

Let us first give a simple example of how to find
quantum geometries. Following this, it will be easier to discuss
the more general approach.
In the original calculation \cite{mye1}, Myers considered $N$
$D$0-branes in a constant four-form Ramond-Ramond (RR) field
strength background of the form
$F^{(4)}_{t123}=-2f\epsilon_{123}$. The relevant $D$0-brane action
is given by \bea S_{Do}=\mu_0\lambda^2\int{ dt\;{\rm
Tr}\left({1\over 2}\partial_t\Phi^i\partial_t\Phi_i -
{1\over4}[\Phi^i,\Phi^j][\Phi_j,\Phi_i] + {i\over
3}F^{(4)}_{tijk}\Phi^i\Phi^j\Phi^k\right)}, \eea where $\Phi^i$
are complex matrix-valued scalars. 
They represent the nine directions $i=1,2,..,9$
transverse to the $D$0-brane. $\mu_0$ is the charge of the
$D$0-brane and $\lambda=2\pi\alpha'$ ($\alpha'$
has dimension $({\rm length})^2$). 
For static configurations, the
kinetic term vanishes and center-of-mass degrees of freedom
decouple. Hence, variation of the action gives the following
polynomial equation in $\Phi$, \bea
[\Phi^j,[\Phi^j,\Phi^i]]+if\epsilon_{ijk}[\Phi^j,\Phi^k]=0 \eea
where indices are contracted with flat Euclidean metric.

This is solved by setting to zero all but the first three scalars
$\Phi^i$ ($i=(1,2,3)$) which are replaced by Lie algebra
generators $T^i$ (in this case angular momentum operators) times a
scalar $r$. Then the above equation becomes an equation for $r$,
\be r-f=0. \ee Therefore, three of the scalars $\Phi^i$ are of the
form $\Phi^i=fT^i$. The non-commutative geometry appears since
these $\Phi$ also correspond to the first three cartesian
coordinates transverse to the brane.

However, as explained in appendix B, the above equations do not
fully define the quantum geometry. It is still necessary to fix
the representation of $T^i$ as different representations give
different solutions and topologies. Each of these solutions has a
characteristic energy $E$, that is a function of the quadratic Casimir
as given by
\bea E=-{\mu_0\lambda^2f^2\over
6}\sum^{3}_{i=1}\hbox{Tr}[\Phi^i\Phi^i]. \eea Given a fixed size
of the representation $N$, the irreducible representation
corresponds to the lower bound (strictly speaking this is the
fuzzy sphere), while the reducible representations have higher
energies corresponding to more complicated topologies of the
direct sums of fuzzy spheres. An important characteristic of the
above construction is that in the large $N$ limit the algebra of
``functions'' defined by these solutions becomes the algebra of
functions on the classical manifold (see appendix B).

This example contains all of the ingredients that define the
process of finding the non-commutative solutions. Generalizations
of this program have appeared, but the underlying structure is the
same. These solutions are usually called ``fuzzy spaces'',
although not all of them are properly well-defined, the best-known
such example being the so-called fuzzy four-sphere\footnote{In
this case it is known that the algebra of functions defined on the
``fuzzy $S^4$" does not close and some extra structure is needed
to properly define the quantum geometry 
\cite{ram1,Balachandran:2002ig}.}

In any case, we can now describe the general picture in terms of
the following basic steps:
\begin{itemize}
\item The starting point is the
non-Abelian action of $N$ $Dp$-branes in the presence of non-trivial
background and world-volume fields.
\item
This action is expanded in terms of the polynomials of the scalar
fields $\Phi^i$ and their world-volume derivatives, where each
monomial comes with a symmetric and/or skew-symmetric product of
the $\Phi^i$'s which translates into commutator and
anti-commutator expressions. The background fields are being
understood as couplings of the world-volume theory.
\item
The equations of motion therefore involve a set of polynomials
in $\Phi^i$ containing  
their commutators and anti-commutators.
\item
Then, one identifies a
subset of the $\Phi^i$'s with some
elements of a Lie algebra times a function $r$ commuting with the
Lie algebra elements. 
The idea is
that the algebraic structure will take care of the commutators and
anti-commutators while $r$ will solve the remaining
equations (possibly differential equations in the world-volume
variables).
\item Finally, one finds different solutions for the field $r$
 corresponding to each of the different representations of the Lie
algebra generators identified with the $\Phi^i$'s. Each of the
different representations encode different topologies and
geometries.
\end{itemize}

We emphasize two important aspects in the above
program. First, the representations associated with scalars
$\Phi^i$ are not fixed by the equations of motion. Second, the
possibility of topology change is suggested by the natural decay of
higher energy solutions into lower energy ones. This cascade
process has already received some attention in \cite{decay}, the
main result being the discovery of unstable modes that trigger the
topology change. These modes are related to the relative positions
of the different fuzzy manifolds that appear in the 
higher energy solutions: given a solution $A$, we can
always construct another solution (of higher energy and more
complicated topology) by considering larger matrices of two or
more copies of $A$. The different topologies of the above types of
solutions come from the fact that they correspond to reducible
representations.

Nevertheless, it is important to note that this is not the only
way of obtaining topology change. There is also the possibility of
having different quantum geometries defined in the same group
structure, which are not related by the simple
``reducible-irreducible'' relations. For example, $SU(4)$ contains
many different fuzzy geometries among which we have
$CP(3),CP(2)$ and $S^2\times S^2$ (see appendix B).

Regarding these aspects, it is obvious that if the
representation is fixed by the equations of motion, there is no
room for topology change. The fact that we can choose the
representation signals the existence of a degeneracy in the set of
quantum geometries, a symmetry that allows topology change. What
we have seen in this section is that from the point of view of the
$D$-brane physics, the equations of motion only define the group
structure, while energy is (in all of the solutions found until now) 
the only quantity that differentiates between quantum geometries and
therefore ``chooses'' the classical geometry in the large $N$
limit via the decay processes.

Therefore, we have a clear relation between topology change and
the fact that the equations of motion do not fix the
representation of the group. Although all of the solutions that
are currently known share the latter feature, there is no reason to
believe that there are no circumstances in which the equations of
motion fix the representation uniquely and hence preclude topology
changing processes. To investigate this matter we need to
understand in greater depth the definition of fuzzy geometry and
the relations between fuzzy coordinates, representations and
invariants of the algebraic structure.

An intuitive way to understand a fuzzy geometry is to define it by
a deformation of the algebra of functions in classical geometry.
For example, take the functions on the sphere. It is enough
to consider the spherical harmonics. In a classical case there is
an infinite number of these. Now, truncate the basis at a given
angular momentum $j$ by projecting out all the higher angular
momentum modes. The resulting algebra 
(with $*$-product replacing point-wise multiplication)
represents the fuzzy sphere
(see \cite{Madore,Balachandran:2002ig,GPFSP} and references therein).

An important mathematical point is that this algebra  of functions
can be obtained from  the symmetric irreducible representations of
size $2j+1=N$ of $SU(2)$  using coherent-state techniques (see
appendix B). Actually, there exists a precise relation between
Cartesian coordinates in $R^3$ (which defines the embedding of the
sphere in $R^3$) and the $SU(2)$ Lie algebra generators in the
corresponding representation. The latter are the three matrices
$T^i$ appearing in the dielectric effect of Myers (note that ${\rm
Tr}(T^iT^i)=NR^2$, with $R$ equal to the radius of the fuzzy
sphere in a suitable length unit).

Nevertheless, the sphere is a sort of a degenerate case since
skew-symmetric representations are trivial for $SU(2)$ and there
are only two basic tensor invariants, corresponding to the
structure constants and Cartan metric. In constructions related to
higher rank groups like $SU(n)\ (n\ge 3)$, there are more tensor
invariants and the skew-symmetric representations are non-trivial,
giving more interesting structures.

An interesting generalization of the fuzzy sphere with rich enough
algebraic structure is the 2$n$-dimensional fuzzy complex
projective planes (fuzzy $CP(n)$) \cite{ale1}. These quantum
geometries are strongly related to $SU(n)$. It turns out that to
define fuzzy $CP(n)$\footnote{A detailed construction of fuzzy
$CP(n)$ is presented in appendix B.}, the equation \bea
d_{ijk}T^jT^k=w\,T^i, \label{eq:1} \eea is sufficient, where
$d_{ijk}$ is the usual invariant rank three symmetric tensor for
$SU(n)\  (n\ge 3)$, and $w$ depends in particular on the 
representation used. This
equation will play an important role in the next section.

\section{Model with Obstruction to Topology Change}

In this section we show how to find a concrete example where the
$D$-brane equations of motion include an invariant tensorial
equation, (like equation \ref{eq:1}) that determines the
representation and hence enforces an obstruction to topology
change. In doing so we will use (\ref{eq:1}) as a hint
and will search for a simple configuration where fuzzy $CP(2)$
could appear.

In order to do this, consider $N$ $D$1-branes with a constant
world-volume electric field $F_{\tau\sigma}$ (the $F_{[2]}$ in
what follows) 
in the presence of a
five-form RR field strength $F_{ijklm}$
(the $F_{[5]})$),
flat metric, constant
dilaton and zero B-field. The energy density for this system is
obtained by expanding the non-Abelian action proposed by
Myers\footnote{See appendix A for a detail derivation and
conventions.} followed by the Legendre transformation. Here we
show the final form for the energy density, once we have restricted our
study to static and constant configurations: \bea
E(\Phi,F_{[2]},F_{[5]})&=&\mu_1\lambda^2\left[{1\over
4}\Phi^{ij}\Phi_{ji}-
{1\over 4}F^2_{\tau\sigma}\right]+ \nonumber \\
&&+ \mu_1\lambda^4\left[ {1\over 16}(\Phi^{ij}\Phi_{ji})^2
-{1\over 8}\Phi^{ij}\Phi_{jk}\Phi^{kl}\Phi_{li}-
{1\over16}F^2_{\tau\sigma}\Phi^{ij}\Phi_{ji}+\right. \nonumber \\
&&\hspace{4cm}\left. +{1\over
10}\Phi^i\Phi^j\Phi^k\Phi^l\Phi^mF_{mlkji}F_{\tau\sigma}\right].
\label{eq:2} \eea Here we have used the convention
$\Phi^{ij}=[\Phi^i,\Phi^j]$, $i,j... \in [1,2,..,8]$. 
This expression is bounded from below since the term 
proportional to $\Phi_{ij}^2$ is positive and 
grows faster then any other with the size of the representation 
in which $\Phi^i$ is.
We also choose the five-form to be 
\bea F_{ijklm}=-h\;d^{pq}_{\;\;\;[i}f_{\;\;jk}^pf_{\;\;lm]}^p\;,
\label{eq:f5} \eea where ``$h$'' represents the strength of the RR
field. Note that 
up to normalization
this is the only invariant tensor with five
indices in $SU(3)$ ($F_{ijklm}$ is related to ${\rm Tr}(U^{-1}\,d\,U)^5$,
the closed 5-form in $SU(3)$).

To find the extremal points of this potential we will use
configurations such that $\Phi^i$ is proportional to the
generator $T^i$ of $SU(3)$, i.e. \bea \Phi^i=\rho T^i. \label{eq:t}
\eea The detailed form of the equations of motion can be found in
Appendix A. These equations are complicated and do not shed extra
light on the discussion. For our purposes it is enough to show the
general structure. From equation (\ref{eq:a1}), we get \bea
T^i=w\,d_{\,ijk}T^j\,T^k, \eea where $w$ is a function of
$N,h,F_{[2]}$ and $\rho$. These equations can only be solved if
the $SU(3)$ generators are in the specific representation of
appendix B. Then, once this representation is chosen, the above
expression becomes an algebraic equation defining $\rho$ as a
function of $F_{\tau\sigma},h$ and $N$. Also, all the matrix products in
the field equations simplify. To study the stability of these
solutions in terms of the variable $\rho$, it is sufficient to
substitute the ansatz back into the expression for the energy
density \bea E =\mu_1\lambda^2N\Bigg[-\hbox{${1\over
4}$}F_{[2]}^2+ \hbox{${3\over16}$}(4-\lambda^2F_{\tau\sigma}^2)c_2\rho^4
+\hbox{${9\lambda^2\over 32}$}(c_2-1)c_2\rho^8+\;\;\;\;
\;\;\;\;\;\; \nonumber \\
-\; \hbox{${\lambda^2\over
40}$}F_{\tau\sigma}h(2n+3)c_2\rho^5\Bigg], \eea where
$N=(n+1)(n+2)/2$, $c_2=(n^2+3n)/3$ is the quadratic Casimir of the
symmetric representation of $T^i$ and $n$ is an arbitrary positive
integer.

\begin{figure}
\hspace{0.22in}
\includegraphics[angle=-90,scale=0.57]{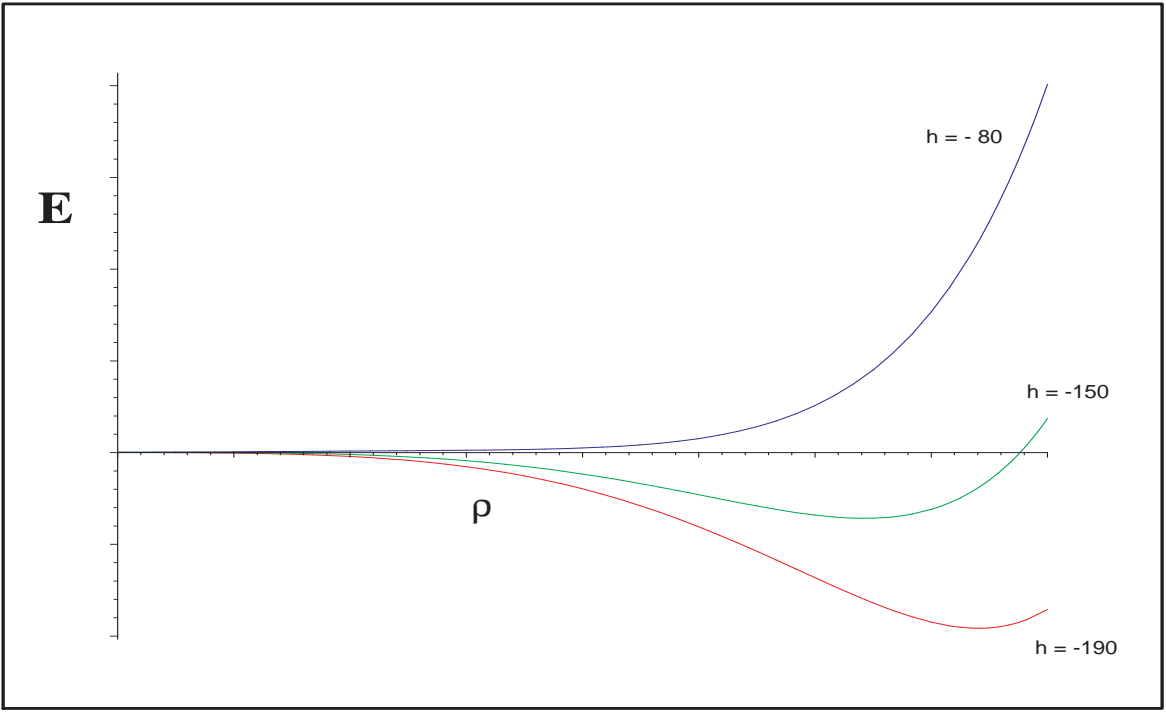} \vspace{-0.0in}
\caption{{\it Energy density for the $D$-string with non-trivial
world-volume electric field $F_{[2]}$ and background RR 5-form
field strength $h$. In the figure we show the potential as a
function of the radius $\rho$ for three values of $h$,
$F_{\tau\sigma}=\sqrt{2}$ and $ \lambda=1$. (See
Appendix A for definitions).}} \label{fig1}
\end{figure}

This energy density has a global minimum at some value of
$\rho$ that we will call $\rho_0$. It depends on the value of the
electric field $F_{[2]}$, the strength of the background RR field
$h$ and $N$. Figure (\ref{fig1}) shows the plot of this $E$.
In particular, if we fix $F_{\tau\sigma}=\sqrt{2}$ and
$\lambda=1$, then the expression for $\rho_0$ simplifies giving
\bea \rho_0=\left({h(2n+3)\over 9\sqrt{2}(c_2-1)}\right)^{1/3}.
\eea

$CP(2)$ is embedded in $S^7$ (see Appendix B). The physical radius
of this seven-sphere corresponds in the fuzzy case to \bea
R=\lambda\sqrt{Tr(\Phi^i\Phi_i)/N}=\lambda\sqrt{c_2}\rho. \eea
Solving for $\rho$ in this example, we get \be
R=\left({h(2n+3)c_2^{3/2}\over 9\sqrt{2}(c_2-1)}\right)^{1/3}. \ee
that in the large $N$ limit gives \be R\sim \hbox{${\sqrt{2}h\over
27}$}n^{2/3}. \ee In the above expression we can see how the
radius increases with the number of $D$1-branes and the strength
of the RR field. Also, note that in order to achieve this effect
we needed a nontrivial electric field on the $D$1-brane, so that
we have fundamental strings diluted into the $D$1-brane in our
model.

Therefore, we have found new type of solutions corresponding to
$N$ $D$1-branes forming a fuzzy $CP(2)$, with an obstruction for
topology change. In the large $N$ limit this configuration goes
over a $D$5-brane with topology $R^{(1,1)}\times CP(2)$, where 
$R^{(1,1)}$ comes from the worldsheet expanded by the original $N$ $D$1-branes.

In fact, we can check this correspondence by looking at the
different couplings of the non-abelian $D$1-branes to various RR
fields. For example, 
consider the coupling to the RR 6-form potential $C^{[6]}$, 
\bea {\mu_1\over2}\int{{\rm
Str}\left(P[\lambda^2(i_\Phi i_\Phi)^2C^{[6]}]\right)}. \eea 
where $P$ is the pull-back from space-time to the world-volumme of the 
$D$1-brane, and the symbol $i_\Phi$ is a non-abelian generalization 
of the interior product with the coordinates $\Phi^i$ (see appendix A for 
more definitions and notation). 

Using the fact that $C^{[6]}$ 
must have support on the fuzzy $CP(2)$ we
write 
\bea C^{[6]}=C_{\tau\sigma ijkl}\propto C_{\tau\sigma1\,2\,3\,4}\;
d^{pq}_{\;\;\;[i}f_{\;\;jk}^pf_{\;\;lm]}^p {x^m\over r} \eea 
where the indices $(1,2,3,4)$ stand for the four real directions on the $CP(2)$
and $\alpha=\hbox{${16\over\sqrt{27}}$}$. Hence the Chern-Simons
term (see Appendix A) gives, \be {\mu_1\alpha\over2}\int{d\tau d\sigma {\rm
Str}\left(\lambda^3C_{\tau\sigma1\,2\,3\,4}
{\Phi^i\Phi^j\Phi^k\Phi^l\Phi^m\over R}
d^{pq}_{\;\;\;[i}f_{\;\;jk}^pf_{\;\;lm]}^p \right)}. \ee
Therefore, after using the fuzzy solution (\ref{eq:t}) with
$\mu_1=4\pi^2\mu_5\lambda^2$ and volume of $CP(2)$ equal to
$8\pi^2$ we get, 
\be {\mu_5\alpha N(2n+3)\over16\,c_2^{3/2}}
\int{d\tau d\sigma\;(8\pi^2)\;R^4C_{\tau\sigma1\,2\,3\,4}}. 
\ee 
which in the limit of large $N$ takes the form of
the coupling of a single $D$5-brane \footnote{where the
world-volume of the $D$5-brane is taken along the $(\tau,\sigma)$
and $CP(2)$ directions, and we average over $CP(2)$.}, \be
\lim_{n\rightarrow \infty} {\alpha N(2n+3)\over 16\,c_2^{3/2}}= 1.
\ee There exist other types of examples of non-Abelian $D$-branes
forming higher dimensional $D$-branes, where the resulting
geometry is a fuzzy CP(2) manifold. These cases however, are
different in nature from the one presented here since the
resulting fuzzy geometry is determined by the lowest energy
condition and not by the equations of motion. The dual picture
corresponding to $D$-branes with CP(2) topology has already been
studied, we refer the reader to \cite{tri1} for further
information.

\section{Summary}

In this article we have studied the relation between Lie group
representations, fuzzy geometries and topology change in matrix
models appearing in $M$-theory (i.e non-abelian $D$-branes,
Matrix string theory and M(atrix) theory). 
There are two different
types of transitions that lead to the topology
change: The first type is related to reducible
representations of the Lie group, where a cascade from
reducible representations to an irreducible ones  
does not change the type of the irreducible representation
(i.e the type of coset of the Lie group)
that appear in the decomposition of the 
the original representation: i.e these are the 
transitions of the type $S^2\times S^2 \times \cdots S^2\rightarrow S^2$.
and 
are due to the
difference in energy between these solutions. Recall that
reducible representations are direct sums of irreducible
representations, where each irreducible representation gives a
fuzzy manifold. The second type of topology change is related
to a transition between different types of cosets that can be defined in a
given Lie group (i.e these are transitions of the type
$CP(2)\rightarrow S^2$, etc.). 
Again, these solutions also have different
energies which triggers the topology change. Note, however, that these
cosets define different fuzzy geometries found within the Lie
group.

These topology changing processes are consequences of the fact that
the generic equations of motion do not discriminate between
representations or cosets of the Lie group. There is a degeneracy
between the solutions that translates into topology changing
processes.

By introducing higher rank Lie groups and by turning on more
background and world-volume fields in the effective $D$-brane
action, we were able to find an obstruction to the 
topology change involving transitions between different cosets
(e.g. $S^2\rightarrow CP(2), etc.$)
corresponding to different types of IRR's present in the
decomposition of the solution.
Basically, we found that the corresponding
equation of motion determined the irreducible representation of the
matrix-valued scalars. This is equivalent to fixing the type of coset of
the Lie group, therefore ruling out this type of transition.

These new solutions correspond to $D$1-branes forming $D$5-branes
with topology $R^{(1,1)}\times CP(2)$.

\acknowledgments

We thank V.P. Nair for useful discussions. The work
of PJS was supported in part by NSF grant PHY-0098747 to Syracuse
University and by funds from Syracuse University. GA and APB were
supported in part by the DOE and NSF under contract numbers
DE-FG02-85ER40231 and INT-9908763 respectively.

\appendix
\section{Non-Abelian $D$-brane action}

In the following, we define the conventions used for the $D$-brane
action. We borrow almost all of the conventions from Myers
\cite{mye1,sil1}.

Our starting point is the low energy action for $N$ $D$-strings with
non-trivial world-volume electric field $F^{[2]}$. 
In the background we have a trivial dilaton $\phi$, 
a flat metric $G$ and zero B-field. For the Ramond sector we include 
a 4-form potential $C^{[4]}$. The action of the $N$-D1-branes has two 
separated terms, the Born-Infeld and the Chern-Simon actions. 
The Born-Infeld action is given by \bea
S_{BI}=-T_1 \int d\tau d\sigma \, {\rm STr}\left( e^{-\phi}
\sqrt{-\left( P\left[G_{ab}+G_{ai}(Q^{-1}-\delta)^{ij}G_{jb}
\right]+\lambda F_{ab} \right) \, det(Q^i{}_j)} \right),
\label{eq:3} \eea with \be Q^i{}_j\equiv\delta^i{}_j +
i\lambda\,[\Phi^i,\Phi^k]\,G_{kj}\ , \ee and the Chern-Simons action 
is 
 \bea S_{CS}=\mu_1\int d\tau
d\sigma \;{\rm STr}\left( P\left[ e^{i\lambda\,i_\Phi
i_\Phi}(C^{(4)}) \right] \, e^{\lambda F}\right)\ , \eea
where below we explain all the conventions:
\begin{itemize}
\item{}
Indices to be pulled-back to the world-volume (see below) have
been labelled by $(a,b,..)$. Space-time coordinates are labelled by the 
indices $(A,B,..)$. The index $(i,j,...)$ labels only directions 
perpendicular to the $D$1-brane.
\item{}
The parameter $\lambda$ is equal to $2\pi l_s^2$, where $l_s$ is the 
string length scale and $T_1$ is the D1-brane tension.
\item{}
The center-of-mass degrees of freedom do not decouple, but it will
not be relevant for our discussion as we will consider static
configurations independent of the space-like world-volume
direction. The fields $\Phi^i$ thus take values in the adjoint
representation of $SU(N)$. As a result, the fields satisfy ${\rm Tr}\,
\Phi^i=0$ and form a non-abelian generalization of the coordinates
specifying the displacement of the branes from the center of mass.
These coordinates have been normalized to have dimensions of
$(length)^{-1}$ multiplied by $\lambda^{-1}$.
\item{}
$P$ stands for the non-abelian ``pullback'' of
various covariant tensors to the world-volume of the $D$1-brane.
We will use the static gauge $x^0=\tau,x^1=\sigma,x^i = \lambda
\Phi^i$ for a coordinate $x$ with origin at the $D$1-brane
center-of-mass. 
\item{}
The symbol $STr$ will be used to denote a trace over the $SU(N)$ 
indices with a complete symmetrization over the non-abelian objects 
in each term. The symbol $i_\Phi$ is a non-abelian generalization 
of the interior product with the coordinates $\Phi^i$, for example 
given a 2-form RR potential $C^{[2]}$,
\be i_\Phi \left(\frac{1}{2}C_{AB}dX^AdX^A\right) = \Phi^iC_{iB}dX^B. 
\ee
\end{itemize}

If we restrict our study to static configurations involving eight
nontrivial scalars $\Phi^i$, ${i=1,..,8}$, the world-volume field
strength $F^{[2]}=F_{\tau\sigma}\;d\tau \wedge d\sigma$ and the
the RR field strength $F_{[5]}$, the above action (Born-Infeld plus
Chern-Simons) gives the following Lagrangian density: \bea
&&L(\Phi,F^{[2]},F_{[5]})=-\mu_1\lambda^2\;{\rm STr}\left[{1\over
2}(\partial\Phi)^2+
{1\over 4}\Phi^{ij}\Phi_{ji}+{1\over 4}F^2_{[2]}\right]+ \nonumber \\
&&-\mu_1\lambda^4\;{\rm STr}\left[ {1\over
4}\partial\Phi^i\partial\Phi_i\Phi^{jk}\Phi_{kj}- {1\over
2}\partial\Phi^i\partial \Phi^j\Phi_{jk}\Phi^k_{\;\;i}-
{1\over 8}\Phi^{ij}\Phi_{jk}\Phi^{kl}\Phi_{li}+\right. \nonumber \\
&&\hspace{2cm}\left. +{1\over
16}(\Phi^{ij}\Phi_{ji})^2+{1\over16}F^2_{[2]}\Phi^{ij}\Phi_{ji}+
{1\over
10}\Phi^i\Phi^j\Phi^k\Phi^l\Phi^mF_{mlkji}F_{\tau\sigma}\right],
\label{eq:a5} \eea  
where $\Phi^{ij}\equiv [\Phi^i,\Phi^j]$.

The equation of motion obtained from the variation of $\Phi^i$ in
(\ref{eq:a5}) gives \bea (1+{1\over
4}\lambda^2F^2)[\Phi^j,\Phi_{ji}]+{1\over2}\lambda^2
[\Phi^j,\Phi_{ji}(\Phi^{kl}\Phi_{lk})]-{1\over2}\lambda^2[\Phi^j,
(\Phi_{jk}\Phi^{kl}\Phi_{li}-\Phi_{ik}\Phi^{kl}\Phi_{lj})]&& \nonumber \\
+{1\over2}\lambda^2F_{\tau\sigma}\{\Phi^{jk},\Phi^{lm}\}F_{ijklm}=0.
\,\;\;\;\;\;\;&&
\eea 
The $U(1)$ part of this equation involves the coupling to the
world-volume electric field and can be solved trivially using the
ansatz of constant electric field. That leaves only the $SU(N)$
part. Using the ansatze (\ref{eq:f5}) and (\ref{eq:t}), we then get
the  equation \bea \left[(1+\hbox{${\lambda^2F^2\over
4}$})3\rho^3+\hbox{${9\over4}$}(c_2-1)\lambda^2\rho^7\right]T^i
-\left[\hbox{${3\over4}$}\lambda^2F_{\tau\sigma}h\rho^4\right]d^{ijk}T_jT_k=0,
\label{eq:a1} \eea where $c_2$ is the quadratic Casimir operator
constructed from $T^i$, and we have used the fact that the quartic
Casimir operator of $SU(3)$ is proportional to the square of the
quadratic Casimir $c_2$ \cite{oku1}.

\section{Fuzzy Geometry and $CP(n)$}

In this appendix we will give a brief description of some of the
ideas used in non-commutative geometry that are relevant to this
paper and derive all the necessary equations used in the previous
chapters. This introduction will be sketchy, since being
an active field there are many new papers that appear almost
daily. We refer any interested reader to one of the several
reviews available in the literature
\cite{Connes,Madore,Balachandran:2002ig,Douglas:2001ba}.

The field of non-commutative geometry is not new
\cite{Snyder:1946qz}. Recently, it has been given much attention
due to the appearance of non-commutative effects in the low energy
effective physics of $D$-branes - the very effect we are studying
here. However, one does not have to think of non-commutativity
as being only the effective description of the theory.
Assuming the fundamental structure of space-time to be that of
some non-commutative algebra, one can try to derive
corresponding consequences for the large-distance (low energy)
physics. By its very nature this point of view leads to mixture of
the ``gravitational'' and field theory degrees of freedom.
\cite{Madore,Balachandran:2002ig,Connes}.

There is
another, more pragmatic reason for introducing  non-commutative
manifolds into physics. It is the ever-present need for
regularization of quantum field theories (QFT's). The usual,
cut-off or
lattice regularizations are very successful in many
numerical aspects of the problem, but are usually associated with
breaking of space-time symmetries of the underlying theory.
They produce such unpleasant effects as fermion doubling
or loss of general covariance in intermediate
computations.
However, when introducing the non-commutativity between
coordinates,
one can at times include the
symmetry algebra
one seeks to preserve as part of the algebra generated by coordinates. 
In this case, if the Lagrangian is invariant
under the symmetry transformations, the corresponding
noncommutative theory will have the symmetry preserved.
(The best example is the fuzzy sphere \cite{FuzzyS2,Balachandran:2002ig}).

Let us now describe one particular procedure of obtaining a
non-commutative manifold. We start by promoting the coordinates
$\xi_i$ of the system to become operators $\hat\xi_i$ satisfying
the non-commutative algebra
\begin{eqnarray}
\xi_i\rightarrow \hat\xi_i\ \ \ \ \left[
\hat\xi_i,\hat\xi_j\right]=C(\hat\xi)_{ij}, \label{oper}
\end{eqnarray}
where $C(\xi)_{ij}$ is a skew-symmetric function of $\xi$
(with definite ordering).
Then, one looks for all possible matrix representations
of this algebra. Each representation is
realized by a set of  operators acting on a  Hilbert space.

At this point we still do not know what "manifold" we are talking
about - the reason being that the very notion of a "point" has
disappeared: there are only matrices now. In general,
properties of the manifold can be encoded in the different
characteristics of the algebra of functions defined over it. If
one wants to know which functions one can get
in the commutative limit
(i.e. limit when $C(\hat\xi)_{ij}\rightarrow 0$),
it is convenient
to introduce
the notions of coherent states (CS) and the diagonal coherent state
representation \cite{CS}. The coherent state of our interest
is obtained by acting with the group element $g$
in one particular representation $g\rightarrow T(g)$ on a highest weight
vector $|\mu\rangle$ of the Hilbert space associated with
the representation:
\begin{eqnarray}
|g\rangle \equiv T(g)|\mu\rangle.
\end{eqnarray}
For any operator $\hat O$ in the Hilbert space one can
compute the so-called {\it symbol} of the operator, defined as
the diagonal matrix element over the  different coherent states:
\begin{eqnarray}
O(g)=\langle g|\hat O|g\rangle.
\end{eqnarray}
Due to the over-completeness of the generic set of coherent states,
the symbol (B.3) of the operator contains information about
{\it any} matrix element of this operator and determines
the operator.
These symbols give the functions $O$ on the
manifold. Suppose that a highest weight vector of some representation
transforms by a singlet representation of a subgroup $H$ of $G$:
\begin{eqnarray}
H\subset G,\ \ \ \forall h \in H: \ \ T(h)|\mu\rangle =
e^{i\alpha(h)}|\mu\rangle.
\end{eqnarray}
So $|\mu\rangle$  changes only by a phase $e^{i\alpha(h)}$
($\alpha(h)$ is  real).
Due to this feature one can see that
\begin{eqnarray}
O(g)&=&\langle g|\hat O|g\rangle=
\langle\mu| T^{\dagger}(g)|\hat O |T(g)|\mu\rangle
\nonumber
=\langle\mu| T^{\dagger}(h) T^{\dagger}(g)|\hat O |T(g)T(h)|\mu\rangle =
\\
&=&\langle\mu| T^{\dagger}(gh)|\hat O |T(gh)|\mu\rangle
\Leftrightarrow O(g)=O(gh),\ \ \ \forall h \in H.
\end{eqnarray}
This invariance property means that the functions on $G$ that one
gets this way are identical to functions on
the coset $G/H$. {\it This shows that
the non-commutative  manifold is defined not only by the right
hand side of (\ref{oper}), but also by the representation
chosen}.


After doing all of the above, one can construct the ``field
theory'' on this manifold. Values of the fields become matrices
and all the integrals over volume become traces in this Hilbert
space $\cal H$: 
$\int V(\Phi) \rightarrow {\rm Tr}_{\cal H}(V(\hat \Phi))$. {\it A
priori}, the dimension of this Hilbert space (representation) can
be  infinite, i.e. traces can include infinite summations. Thus
from the point of view of regularization, the main goal is to
obtain suitable manifolds associated with
finite-dimensional Hilbert spaces. Generally this cannot be done, 
but there is a class of manifolds for which this is
virtually guaranteed - these are the co-adjoint orbits of compact
Lie groups \cite{Kirillov}. This means that the right hand side of
(\ref{oper}) should be $C(\hat\xi)_{ij}=i\,f_{ijk}\,\hat\xi_k$, where
$f_{ijk}$ is the appropriate (e.g. $SU(N)$) structure constants.

Let us now give several examples which have been studied in the
literature using some of the ideas described above:
\begin{itemize}
\item
"Moyal plane". In this case one has only two dimensions $x_1,x_2$,
and $C_{ij}$ is just $\epsilon_{ij}$. This is the well-known
harmonic oscillator ($h.o.$) algebra and the Hilbert space is the
infinite-dimensional set spanned by linear combinations of of $h.o.$
excited states, $|n\rangle$.
\item
The "fuzzy" sphere $S^2=CP(1)$. It is the coset
$SU(2)/U(1)$ and is the orbit of say the $\sigma_3$ generator 
under the adjoint action of
$SU(2)$. The corresponding algebra is just the usual angular
momentum one, with $C_{ij}=i\,\epsilon_{ijk}\xi_k$. All the irreducible
representations (IRR's) can be obtained from
symmetric products of the fundamental
representation and can be labelled by  $j \in Z^+/2$. All the
fields become $(2j+1)\times(2j+1)$ matrices and all the traces
become finite (there are only $2j+1$ terms in each sum.) One can
then put field theory  on this fuzzy manifold and obtain, for example,
an explicit expression for the path integral which is finite
dimensional \cite{KGP,Balachandran:2002ig}. It is quite interesting that upon
introducing fermions in the model, there are also arguments why
this construction avoids the famous fermion doubling problem (see
\cite{GPFSP}, the first three papers of \cite{FuzzyS2}
and \cite{Balachandran:2002ig}).
\item $CP(2)=SU(3)/U(2)$. This coset is the $SU(3)$ orbit
of the ``hypercharge'':$Y=diag(1,1,-2)$ under the adjoint action. 
The corresponding
representations are totally symmetric products of $3$'s or $3^*$'s.
The corresponding Hilbert space has dimension $(n+1)(n+2)/2$
for any positive integer $n$.
This manifold has been obtained as one of the solutions in
\cite{tri1,Nair:1998bp} and analyzed in \cite{GPCP2,ale1}.

\item $SU(3)/(U(1)\times U(1))$. This is the "other" coset of the
$SU(3)$ which was obtained as one of the possible
solutions in \cite{tri1}.
It is the orbit of the (1,-1,0) generator of $SU(3)$ 
under the adjoint action and
produces representations which have zero hypercharge.

\end{itemize}

For the purposes of this paper, we need one of the $CP(n)$ type
manifolds, namely $CP(2)$. It has been extensively studied  in 
\cite{GPCP2,ale1}. 
Here we will present only the relevant facts along the lines of
Alexanian et al. \cite{ale1}.

The classical, "continuous" $CP(2)$ manifold can be obtained as
one particular coset of $SU(3)$: $CP(2)=SU(3)/U(2)$ (this is
general: $CP(n)=SU(n+1)/U(n)$). What is important for us is that
$CP(2)$ is the adjoint orbit of the hypercharge in $SU(3)$, i.e.
\begin{equation}
t^a\,\xi^a={\rm constant}\times U\,t^8\, U^{-1}, \ \ \ \ U\in
SU(3),\ \  a=1,2,...8,
\label{defcp2}
\end{equation}
repeated indices being summed over.
Here $t^a$'s are generators of $SU(3)$ in the fundamental
representation (where $t^a=\lambda^a/2$, $\lambda$'s being the
Gell-Mann matrices). In this formula $\xi^a$ are coordinates in
$R^8$ and $U$ is an arbitrary $SU(3)$ matrix. This equation
defines $CP(2)$ as a surface in $R^8$. 
With the constant=1 for simplicity, squaring both sides of (\ref{defcp2})
and tracing,
we also get 
\begin{equation}
\xi^i\,\xi^i=2\,{\rm tr}(U\,t^8\, U^{-1}\,U\,t^8\, U^{-1})=
2\,{\rm tr}(t^8\,t^8)=1.
\end{equation}
Therefore $CP(2)\subset S^7$.

Using the property of $t^8$
(or hypercharge) that
\begin{equation}
(t^8)^2 = \frac{1}{6}-\frac{1}{2\sqrt{3}}t^8,
\end{equation}
one can show that the $\xi$'s so defined also satisfy
\begin{equation}
\xi^a\,=\,{\rm constant}\times d^{abc}\,\xi^b\,\xi^c,
\end{equation}
where $d^{abc}$ is the standard totally symmetric traceless invariant $SU(3)$
tensor. The remarkable fact is that this
statement can be reversed, i.e.
\begin{equation}
\xi^a\,=\,{\rm constant}\times d^{abc}\,\xi^b\,\xi^c \ \ \Rightarrow \ \
\xi \in CP(2), \label{deq}
\end{equation}
so that this equation can be used to define $CP(2)$ \cite{ale1}.

In the non-commutative case, the coordinates $\xi^i$ will become
operators $\hat\xi^i$ in an appropriate Hilbert space which satisfy $SU(3)$
commutation relations.
Naively, one may try to use any irreducible representation of
$SU(3)$ to represent the $\xi$'s. However, only the totally
symmetric
ones  produce $CP(2)$
in the commutative limit \cite{ale1}.
This is because only the symmetric representations have highest weight
vectors with the $U(2)$ stability group, which, according to the
discussion above leads to the $SU(3)/U(2)=CP(2)$
coset in the continuous limit.

We will show now that the imposition of the condition (\ref{deq})
as an operator equation
\begin{equation}
\hat\xi^a\,=\,{\rm constant}\times d^{abc}\,\hat\xi^b\,\hat\xi^c, \ \
\label{opeq}
\end{equation}
{\it allows only totally symmetric representations} of $SU(3)$.
First we note that any irreducible
representation of the group $SU(3)$ can be obtained from
the direct product of the totally symmetric product of
fundamental (with generators $t_i$) and the totally symmetric product of
anti-fundamental(with generators $-{t^*}_i$) representations.
Assuming for the
moment that the fundamental representation satisfies (\ref{opeq})
for some value of the ``constant'', we can immediately see
that by replacing $t_i\rightarrow -t^*_i$, the ``constant'' will have
to change sign as well. Therefore, any representation that has
both fundamental and anti-fundamental components present in its
decomposition cannot satisfy this equation with ${\rm constant} \,\ne\, 0$.
This means that only totally symmetric products of the fundamental
(or anti-fundamental) representations are allowed.

This is exactly what we have to use in the text to obtain
$CP(2)$ as a solution of the equations of motion,
and not a choice of the energy condition.

In order to do an explicit calculation one can use the Schwinger
representation for the generators $\hat\xi^a$. Using three
harmonic oscillators 
with annihilation and creation operators
$a_i, a_i^+,\ i=1,2,3$ where $[a_i,a^+_j]=\delta_{ij}$, one
has $\hat\xi^a=\alpha\,a^+_i(t^a)_{ij}a_j$, where $\alpha$ is some
constant to be determined later.
$\hat\xi^a$ acts on the Fock space with basis
$\prod(a^+_i)^{n_i}|0\rangle=|n_1,n_2,n_3\rangle$ which are
symmetric under interchange of $a^+_i$'s. As
$[\hat\xi^a,a^+_j]=\alpha\,a^+_j(t^a)_{ji}$, we thus get only symmetric
products of the fundamental representation.
For a similar construction for anti-fundamental representation,
we must start from $\hat\xi^a=-\beta\,b^+_i{(t^a)}^*_{ij}b_j$,
where $b_i,b^+_j$ are also bosonic oscillators (commuting with
$a_k,a^+_k$) and $\beta$ is a constant.
Now the operators $\sim a^+\,t^a\,a$
act irreducibly on the subspace spanned by
$|n_1,n_2,n_3\rangle$, with $n_1+n_2+n_3=n$ held fixed, $n$
being an arbitrary positive integer. The dimension $N$ of this
Hilbert space is easily obtained as
$N = (n+1)(n+2)/2$. After a straightforward but somewhat tedious
computation, (see \cite{ale1} for details), one then gets,
\begin{equation}
d_{abc}\hat\xi_a\hat\xi_b=2t^a_{\alpha\beta}\left(
\frac{n}{6}+\frac{1}{4}\right)a^\dagger_{\alpha}a_{\beta}=
\frac{\left(\frac{n}{3}+\frac{1}{2}\right)}{\sqrt{\frac{1}{3}n^2+n}}\hat\xi_a.
\label{cubic2}
\end{equation}
Let us choose the value of $\alpha$ so that
$\sum_i\hat\xi_i\hat\xi_i=I$, the identity operator.
Using the value $\frac{1}{3}n^2+n$
of the the quadratic Casimir in the
symmetric representation
$c_2$ one can see that $\alpha=1/c_2^{1/2}$,
so $\alpha=1/\sqrt{n^2/3+n}$.


\end{document}